\def\half{{\textstyle\frac{1}{2}}}

\documentclass{ws-ijmpa}
\usepackage[super,compress]{cite}
\usepackage{graphicx}
\usepackage{grffile}
\usepackage{amsmath}

\usepackage{hyperref} 
\DeclareUrlCommand\doi{\def\UrlLeft##1\UrlRight{DOI:\nobreakspace\href{http://dx.doi.org/##1}{##1}}\urlstyle{rm}} 

\def\draftnote{}
\let\trimmarks%

\begin{document}
\markboth{Martin Oral, Ond\v{r}ej \v{C}\'{i}p, Luk\'{a}\v{s} Slodi\v{c}ka}{Simulation of Motion of Many Ions in a Linear Paul Trap}

%


%

\title{Simulation of Motion of Many Ions in a Linear Paul Trap}

\author{Martin Oral}
\address{Institute of Scientific Instruments of the Czech Academy Of Sciences\\
Kr\'alovopolsk\'a 147, 612 64 Brno, Czech Republic\\
oral@isibrno.cz}

\author{Ond\v{r}ej \v{C}\'{i}p}
\address{Institute of Scientific Instruments of the Czech Academy Of Sciences\\
Kr\'alovopolsk\'a 147, 612 64 Brno, Czech Republic\\
ocip@isibrno.cz}

\author{Luk\'{a}\v{s} Slodi\v{c}ka}
\address{Department of Optics, Palacky University Olomouc\\
17. listopadu 12, 77146 Olomouc, Czech Republic\\
slodicka@optics.upol.cz}

\maketitle

\begin{history}
	Preprint of an article published in\par
	\href{https://www.worldscientific.com/worldscinet/ijmpa}{International Journal of Modern Physics A}, Vol.~34, \No~36 (2019), 1942003\par
	\doi{10.1142/S0217751X1942003X}\par
	\copyright World Scientific Publishing Company
\end{history}

\begin{abstract}
The quadrupole linear Paul trap is one of the key instruments in building highly stable atomic clocks. However, a frequency reference based on a single trapped ion is limited in stability due to the time needed for the interrogation cycle which cannot be further shortened. A promising strategy is the utilization of multiple trapped ions. The ions of the same kind then repulse each other with the Coulomb force, which is countered by the ponderomotive force of the time depended field in the trap. A few ions form a chain along the axis of a linear Paul trap. Adding more ions (a few tens or hundreds) gives rise to Coulomb crystals. We created an efficient simulation code which calculates the motion of such collections of ions in quasistatic radiofrequency fields of real linear quadrupole traps (including the micromotion). We include a model for laser cooling of the ions. The simulation tool can be used to study the formation and the dynamics of Coulomb crystals under conditions corresponding to various experimental set-ups.

\end{abstract}

\ccode{PACS numbers:}


\section{Introduction}	
Radiofrequency (RF) Paul traps are valuable in the design and in the operation of highly stable optical atomic clocks based on suitable trapped ions \cite{ludlow}.
The traditional setup involves a single ion in an RF trap irradiated with a laser beam.
The frequency of the laser light is then fine-tuned to match that of the photons coming from an electronic transition in the atomic shell.
The stability can be further improved by using heavy atoms (such as Thorium) and the more stable frequencies of their nuclear transitions \cite{Th-trans}, and by setting up so-called Coulomb crystals, to improve the frequency measurement statistics by increasing the number of reference atoms \cite{obsil}.

The ability of time-dependent electric fields to confine charged particles is well understood by means of the ponderomotive approximation \cite{gerlich}.
The approximation can be extended to include the mutual interactions between multiple ions \cite{modes}.
It can describe the formation of {\sl Coulomb crystals}, when the ions assume quasi-stationary positions as a result of the balance between the confining effect of the trap field and the mutual repulsions between the ions.
The Coulomb crystals can only be achieved if the excess kinetic energy of the ions ("temperature") can be dissipated away, such as via various kinds of laser cooling utilizing the Doppler optical frequency shift.

The goal of out contribution is to present a simulation including the micromotion, that is, the calculation of the trajectories directly in the time-dependent field, including their real-time Coulomb repulsion and a damping force representing laser cooling.
That will provide an independent comparison to results from the ponderomotive approximation, namely the formation of the Coulomb crystals and the determination of the residual micromotion.
These characteristics are important in optical atomic clock set-ups.

\section{Equation of Motion and Forces Acting on Ions}
We assume that the trap contains $N$ ions.
Let us index them with $i=1\ ,2\dots\ ,N$, and the associated quantities of each of them are: $\vec{r}_i(t)$ the positional vector at time $t$, $q_i$ the charge, $m_i$ the mass.
The the equation of motion for each of the ions is:
\begin{equation}
\frac{\mathrm{d}\vec{p}_i}{\mathrm{d} t} =  q_i \left( \vec{E}^{\mathrm{ext}}\left( \vec{r}_i(t) \right) + \vec{E}^\mathrm{C}_i \right) + \vec{F}^\mathrm{drag}_i\ ,
\label{eqmot}
\end{equation}
where $\vec{p}$ is the ion's momentum, $\vec{E}^\mathrm{ext}$ is the external electric field generated by the electrode configuration of the trap, $\vec{E}^{C}_i$ is the Coulomb electric field, and $\vec{F}^\mathrm{drag}_i$ is the damping force, such as that introduced by the Doppler cooling.

The trap field typically requires a numerical calculation and a is evaluated using a suitable interpolation formula.
An example of that method is given bellow.

The Coulomb electric field acting on the $i^\mathrm{th}$ particle is:
\begin{equation}
\vec{E}^\mathrm{C}_i = \frac{1}{4\pi\epsilon_0} \sum\limits_{j=1, j\neq i}^{N}q_j\frac{\vec{r}_i-\vec{r}_j}{|\vec{r}_i-\vec{r}_j|^3}\ ,
\label{coulforce}
\end{equation}
where $\epsilon_0$ is the vacuum permittivity.

The drag force $\vec{F}^\mathrm{drag}$ can take different forms.
We adopted a simple linear anisotropic model, respecting the fact that laser cooling with a single laser beam shining in the direction of the unit vector $\vec{u}$ affects the projection of the particle's velocity into that direction:
\begin{equation}
	\vec{F}^\mathrm{drag}_i = - f_i m_i (\vec{v}_i\cdot\vec{u})\vec{u},\\
	\label{dragforce}
\end{equation}
where $f_i$ is the drag coefficient per unit mass, $\vec{v}_i$ is the velocity of the ion, $m_i$ its mass.
The above model corresponds to a bidirectional laser propagation (one laser beam propagates in the direction of $\vec{u}$, another one in the opposite direction).
If there is only one beam in the direction of $\vec{u}$, we set:
\begin{equation}
	\vec{F}^\mathrm{drag}_i = \vec{0}, \quad\mathrm{if}\ \vec{v}_i\cdot\vec{u} > 0\ .
\end{equation}

\section{Calculation and Evaluation of the Electric Field of the Trap}

\begin{figure}
\includegraphics[width=.6\textwidth]{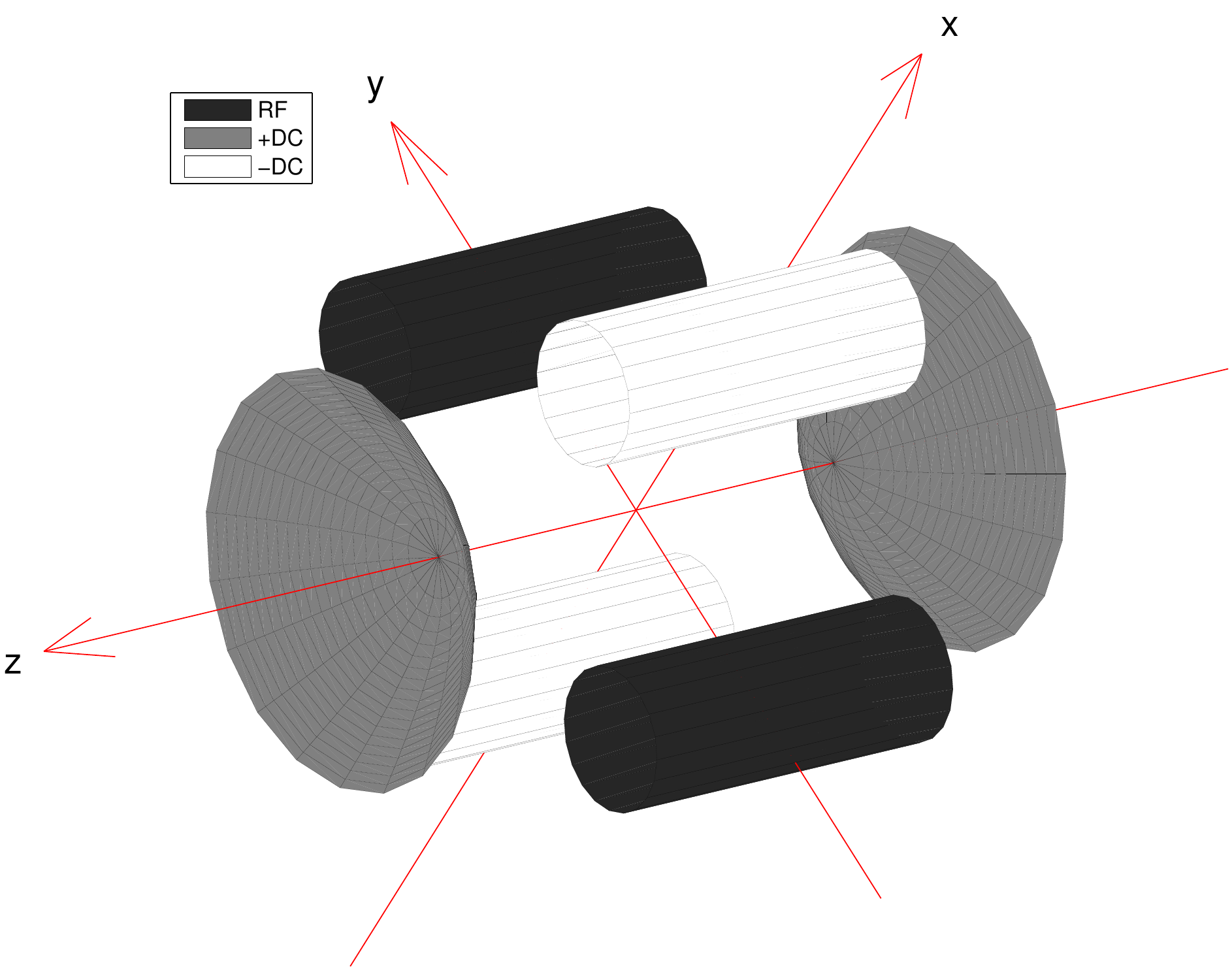}
\caption{Schematic layout of the electrodes in a linear Paul trap.}
\label{trapschema}
\end{figure}

The quadrupole linear Paul trap consists of four quadrupole electrodes and a pair of endcap electrodes.
A schematic drawing is in Fig. \ref{trapschema}.
Both electrodes on the $x$-axis are at the same electric potential $U_x$ relative to the ground.
Likewise, both electrodes on the $y$ axis are at potential $U_y$.
That creates a principal quadrupole field, which needs to be modulated by a periodic time dependent function (sinusoidal) to create the ponderomotive effect which confines charged particles in the trap.
The endcap electrodes are at suitable static potentials that repulse the ions, preventing them from escaping along the $z$-axis.
In the so-called linear traps, the quadrupole electrodes are much longer then their perpendicular distance, the region of confinement in the $z$-direction is relatively long.
Here we assume the endcap electrodes are both at potential $U_c$.

There are two wiring arrangements that are most commonly used in practice. they are summarized in Table \ref{trapwirings}.
\begin{table}
\tbl{Assignment of the trap electrode voltages corresponding to two most common wiring schemes.}
{\begin{tabular}{ll}
\toprule
One electrode pair at AC 			& Both pairs at AC in counterphase \\
\colrule
$U_x:= -U_{DC}$					    & $U_x:=  +\half U_{AC}\cos\omega_{AC}t$ \\
$U_y:= U_{AC}\cos(\omega_{AC}t)$	& $U_y:=  -\half U_{AC}\cos\omega_{AC}t$ \\
$U_c:= 0$							& $U_c:=  U_{DC}$ \\
\botrule
\end{tabular}
\label{trapwirings}}
\end{table}

As the frequency $\omega_{AC}$ is usually in the "radio" range (up to a few tens of MHz), it is possible to adopt quasi-static approximation, that is, the electromagnetic induction is neglected and the associated propagation of the electric field changes is instantaneous.
Then the electric field can be expressed in a form the usual scalar electric potential which is modulated by the time dependent factor of $\cos\omega_{AC}$.

Ideally, the electrodes of the quadrupole should have hyperbolic cross sections, but that is usually not feasible in practice.
The field in real traps then has, besides the dominant quadrupole component, additional higher multiple components, and it must be solved numerically.

\subsection{Decomposition, Symmetries and Antisymetries of the Electric Potential} \label{potdecomp}
To avoid repeated numerical calculations for various values of the voltages $U_{x}$, $U_y$ and $U_{c}$, it is advantageous to decompose the electric potential $\Phi(\vec{r})$ in the trap in the following way:
\begin{equation}
	\Phi(\vec{r})= \left(\frac{U_x+U_y}{2}-U_c\right)P_{++}(\vec{r}) + \frac{U_x-U_y}{2}P_{+-}(\vec{r}) + U_c\ ,
	\label{linkom}
\end{equation}
where the function $P_{++}$ corresponds numerically to the potential distribution generated by electrode voltages $U_x=U_y=1$\,V a $U_c=0$, and the function $P_{+-}$ corresponds to $U_x=-U_y=1$\,V a $U_c=0$.
Then the numerical solution of the "basis" functions $P_{++}$ a $P_{+-}$ only needs to be found once, and a field for any electrode potentials $U_x$  $U_y$ and $U_{c}$, is obtained via the linear combination (\ref{linkom}).

Due to the geometry of the trap (Fig.\ \ref{trapschema}), the basis functions $P_{++}$ a $P_{+-}$ exhibit the following symetries and antisymetries:
\begin{itemize}
\item The planes $x=0$, $y=0$ a $z=0$ are symmetry planes for both $P_{++}$ a $P_{+-}$ (it holds that $P(x,y,z) = P(x,-y,z)$, $P(x,y,z) = P(-x,y,z)$,  $P(x,y,z) = P(x,y,-z)$).
\item The planes $x=y$ a $x=-y$ are symmetry planes for $P_{++}$, and antisymmetry planes for $P_{+-}$ ($P_{++}(x,y,z)=P_{++}(\pm y,\pm x,z)$), $P_{+-}(x,y,z)=-P_{+-}(y,x,z)$ a $P_{+-}(x,y,z)=-P_{+-}(-y,-x,z)$; due to the previous condition, this is only important for the half-plane $x=y, x\geq 0$).
\end{itemize}
These conditions then can be utilized in the numerical models in the form of Dirichlet and Neumann boundary condition by applying them on the respective surfaces.
Then the field needs to be found in only 1/16 of the entire geometry of the trap.
That not only greatly reduces the requirements on the computational resources, but also increases the accuracy, as the symmetry features of the calculated potential distribution are fulfilled exactly.

\subsection{Expansion of the Potential Around the Axis}
Let us define the standard cylindrical coordinates $(r,\varphi,z)$ with the origin in the center of the trap and the axis $z$ coinciding with the main axis of the trap (Fig. \ref{trapschema}), $r$ the distance from the axis, and $\varphi$ the polar angle measured from the $x$ axis.

The electrostatic potential generated by the electrodes at some potentials can be written as a Fourier series in the polar angle $\varphi$, and as a power series expansion in the radial distance $r$. Due to the symmetries and antisymmetries discussed above, the Fourier series will contain the cosine terms only, and each of the basis functions $P_{++}$ and $P_{+-}$ from eq.\ (\ref{linkom}) will only contain some multipole components, that is, multiplicities of $\varphi$.
The form of the expansions can be found in book \cite{hawkesI} .

For $P_{++}$ the lowest multipole is then the rotationally symmetrical one, the next nonzero higher one is the octupole:
\begin{align}
P_{++}(r,\varphi,z)=&\,               p_{0,0}(z)
                    - \frac{1}{4}   p_{0,2}(z)r^2
                    + \frac{1}{64}  p_{0,4}(z)r^4
                    - \frac{1}{2304}p_{0,6}(z)r^6 \nonumber\\
           & + \left( \frac{1}{24}  p_{4,0}(z)  - \frac{1}{480}p_{4,2}(z)r^2 \right) r^4\cos 4\varphi+\mathcal{O}(r^8)
\label{ex++rp}
\end{align}
For $P_{+-}$ the lowest multiple component is the quadrupole, the principal component for this type of trap.
The next nonzero higher multipole is the dodecapole (the multipicity of six):
\begin{align}
P_{+-}(r,\varphi,z)= & \left( \frac{1}{2}p_{2,0}(z)-\frac{1}{24}p_{2,2}(z)r^2+\frac{1}{768}p_{2,4}(z)r^4 \right)r^2 \cos 2\varphi \nonumber\\
               & + \frac{1}{720}p_{6,0}(z)r^6\cos 6\varphi+\mathcal{O}(r^8)
\label{ex+-rp}
\end{align}
For completeness ,let us add that in the free space without free or bound changes the axial functions $p_{\mu,n}$ for each multipole order $\mu$ fulfil the relation:
\begin{equation}
p_{\mu,n}  = p''_{\mu,n-2}\,,\quad n\geq 2 \label{derpodm}
\end{equation}
where the primes denote derivatives with respect to $z$.
The potential around the axis is then completely defined by the functions $p_{\mu,0}$ and their even-order derivatives.

\subsection{Extraction of Multipole Components of the Potential from Numerical Data and Interpolation of the Field} \label{multextr}
The formulas (\ref{ex++rp}) and (\ref{ex+-rp}) can be used to evaluate the basis functions $P_{++}$ and $P_{+-}$, respectively, and in turn the electric potential for any electrode voltages via (\ref{linkom}), if the axial functions $p_{\mu,n}(z)$ are known for a given $z$.

To extract the axial multipole functions $p_{\mu,n}$ we adapted the procedure presented in \cite{cpo8}~.
We first determine their values for a set of discrete values of the $z$-coordinate: $z_k$, $k=1,2,\dots$ .
To do that we arrange the FEM mesh points within a certain radius around the $z$-axis so that they lie in planes perpendicular to it: $z=z_k$.
In each of these planes the values $p_{\mu,n}(z_k)$ are calculated by fitting the expressions (\ref{ex++rp}) or (\ref{ex+-rp}) on the corresponding set of numerical values of $P_{++}(r,\varphi,z_k)$ or $P_{+-}(r,\varphi,z_k)$, respectively.
We do not incorporate the condition (\ref{derpodm}) in this analysis; the coefficients $p_{\mu,n}$ are taken as independent variables in the regression for each $z_k$.

Having obtained the values of the axial functions $p_{\mu,n}$ in isolated points on the $z$-axis, we can evaluate them for any $z$ using a suitable one-dimensional interpolation method.
For that purpose we use a method introduced by Barth, Lencov\'a and Wisselink \cite{slice} using the Taylor expansion
\begin{equation}
p_{\mu,n}(z)=\sum\limits_{l=0} p^{(l)}_{\mu,n}(z_k) \frac{(z-z_k)^l}{l!}
\label{axfuntayl}
\end{equation}
which involves both odd and even derivatives of $p_{\mu,n}$.
In numerical calculations the expansion is truncated after some order $l$; a reasonable choice is to match the highest order of the required derivatives with those in (\ref{ex++rp}) and (\ref{ex+-rp}) plus the next higher odd order.
The values obtained by fitting, $p_{\mu,n+2}$, $p_{\mu,n+4}$ etc., are used in place of the second derivatives.
Each odd-order derivative $p_{\mu,n+1}(z_k)$ is calculated by forming a cubic spline of $p_{\mu,n}$, differentiating it and evaluating it in $z_k$.

The external electric field in eq. (\ref{eqmot}) is $\vec{E}^{\mathrm{ext}}=-\nabla \Phi$ with the potential defined by eq. (\ref{linkom}) using the basis funnctions $P_{++}$ and $P_{+-}$.
The formulas for the the field components are then given by the appropriate differentiations of the expressions (\ref{ex++rp}) and (\ref{ex+-rp}), considering the relations between the derivatives in cylindrical and Carthesian coordinates (since the equation of motion is solved in Carthesian components).

Thus, prior to the application of the interpolation, such as during trajectory integration, each multipole field component is defined by a table of values
\begin{equation}
p_{\mu,0}(z_k),\ p_{\mu,1}(z_k)=p'_{\mu,0}(z_k),\ p_{\mu,2}(z_k),\ p_{\mu,3}(z_k)=p'_{\mu,2}(z_k), \dots
\label{axfunatzk}
\end{equation}
In the numerical evaluation of the field, given a certain $z$, the closest $z_k$ is looked up, and the corresponding set of the functions (\ref{axfunatzk}) is substituted into the expansion (\ref{axfuntayl}), yielding the values $p_{\mu,n}(z)$ at the $z$-coordinate of interest.
They are then used in formulas for the field components resulting from differentiating (\ref{ex++rp}) and (\ref{ex+-rp}), or simply the potential itself is evaluated, if desired.
Depending on the wiring of the trap electrodes, a scaling according to Table \ref{trapwirings}, including the time-dependent modulations, is applied.

\section{Numerical Integration of the Equation of Motion}
We use the ODE solver from the GNU Scientific Library \cite{GSL} to numerically solve the equation of motion (\ref{eqmot}), in which we assume the non-relativistic momentum  $\vec{p}_i=m_i\vec{\dot{r}}_i$ with the dot denoting the derivative with respect to time.
It solves a system of first-order ordinary differential equation.
Since the equation of motion (\ref{eqmot}) is a system of second order equations, we convert it using the substitutions $\vec{v}_i=\vec{\dot{r}}_i$ into twice as many first-order equations.

The GSL ODE solver allows to simply switch between various kinds of integration methods.
Out of them we mostly use the following two Runge-Kutta variants: the classic method of the fourth order with the fifth-order error estimation, and the Prince-Dormand method of the eight order with the ninth-order error estimation.
The error estimation is used in automatic step size control based on per-component relative and absolute tolerances.

We use a straightforward implementation of the Coulomb force (\ref{coulforce}).
We utilize the fact the each term of the sum can be evaluated once and used twice: it is added to the force on the $i^\mathrm{th}$ particle and, with the opposite sign, to the force on the $j^\mathrm{th}$ particle.

\section{Analysis of a Real Trap}

\begin{figure}
\includegraphics[trim=0 240 0 0,clip=true,scale=0.22]{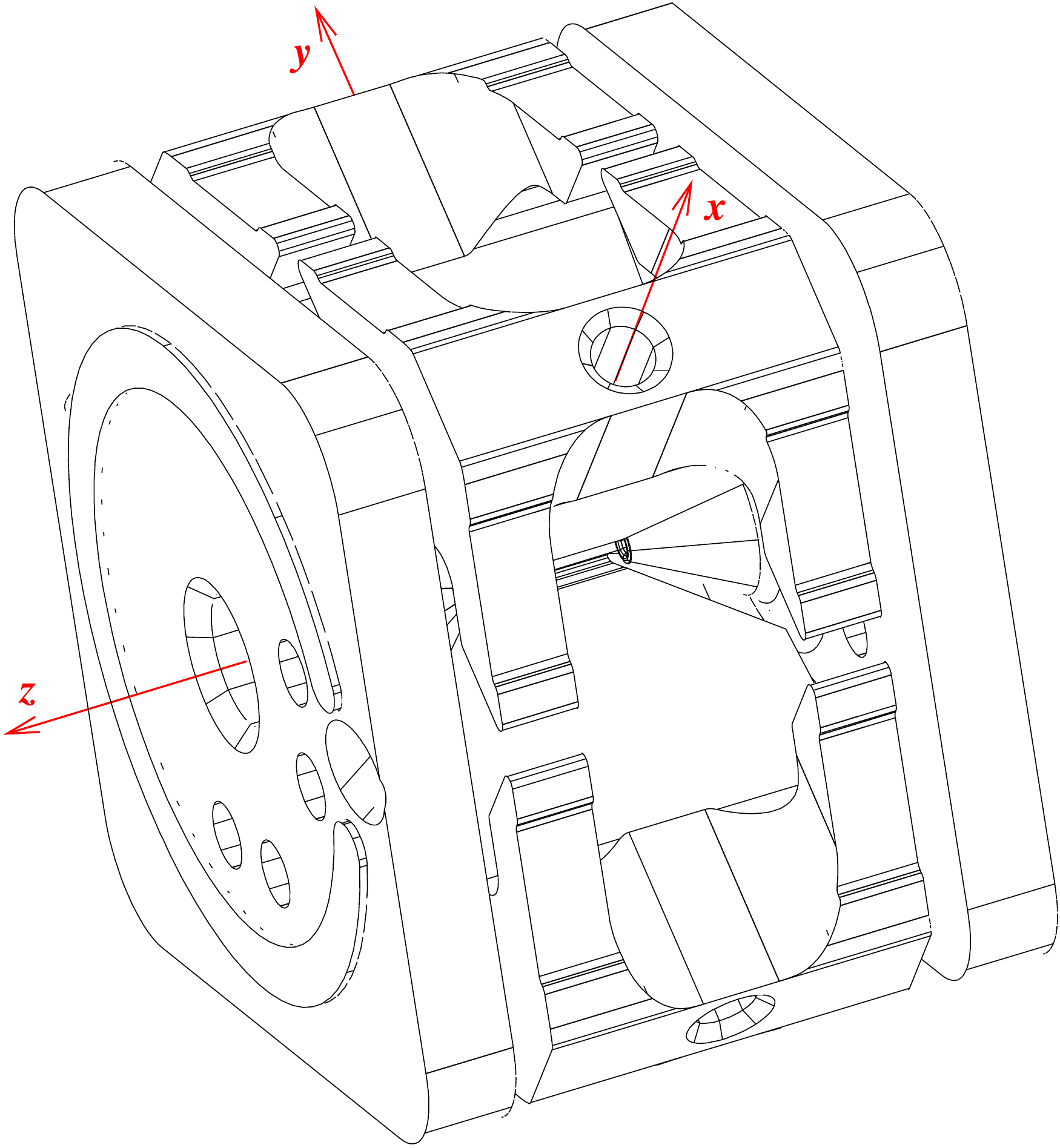}
\includegraphics[scale=0.2]{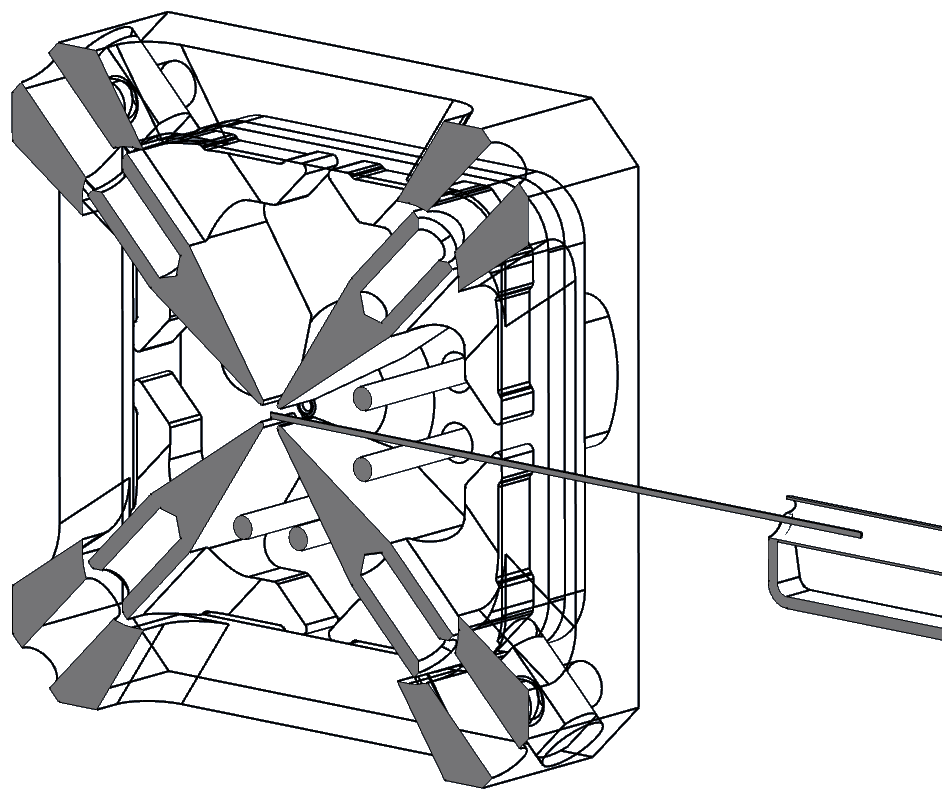}
\includegraphics[scale=0.2]{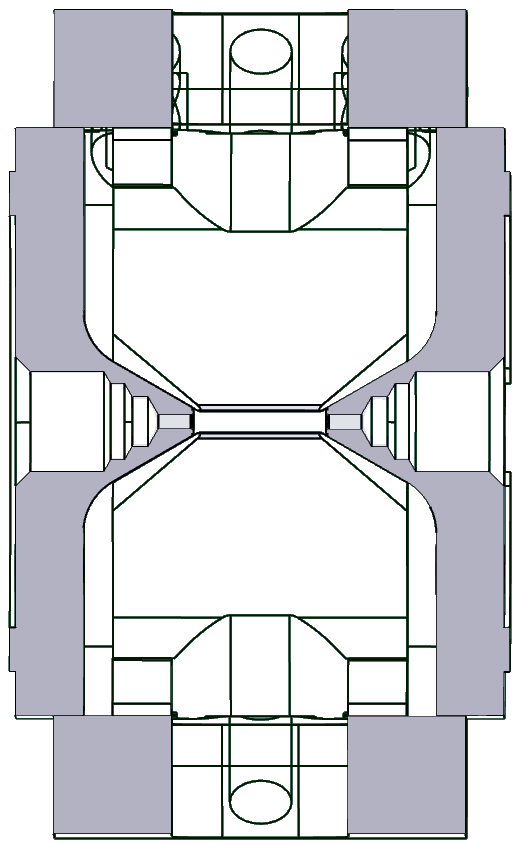}
\caption{Visualisation of the trap. Left: The complete view of the trap with the directions of the Carthesian axes. Center: The section throught the midplane $z=0$, the shaded areas show the profile of the quadrupole electrodes. The structure on the right shows the position of the calcium oven with an indication of the direction at which evaporated calcium atoms arrive at the trap center.
Right: The section through the plane $x=y$, the shaded areas reveal the shape of the endcap electrodes.}
\end{figure}

\subsection{Calculation of the electric field using FEM}
\begin{figure}
\includegraphics[scale=0.45]{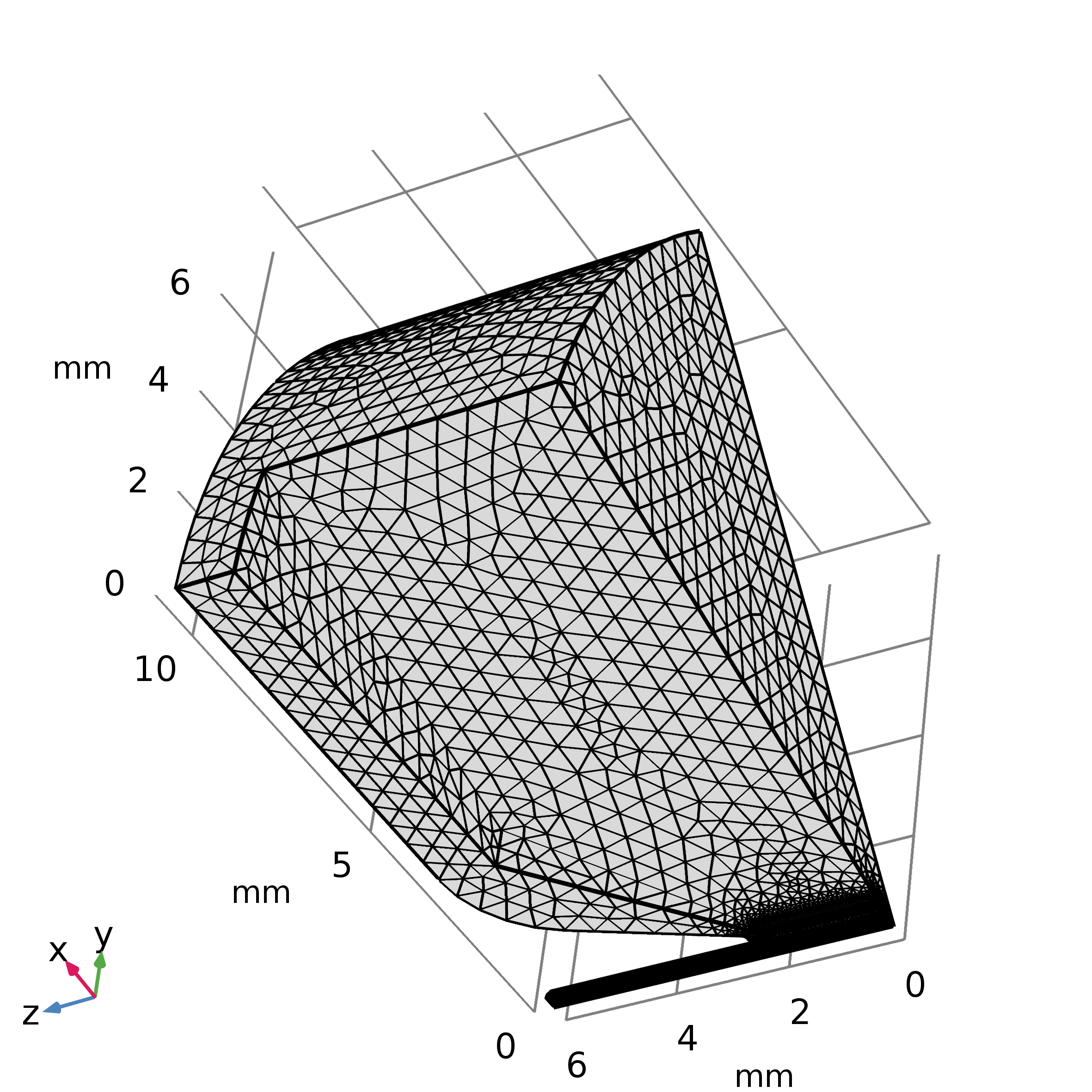}\hspace*{-5mm}
\includegraphics[scale=0.15]{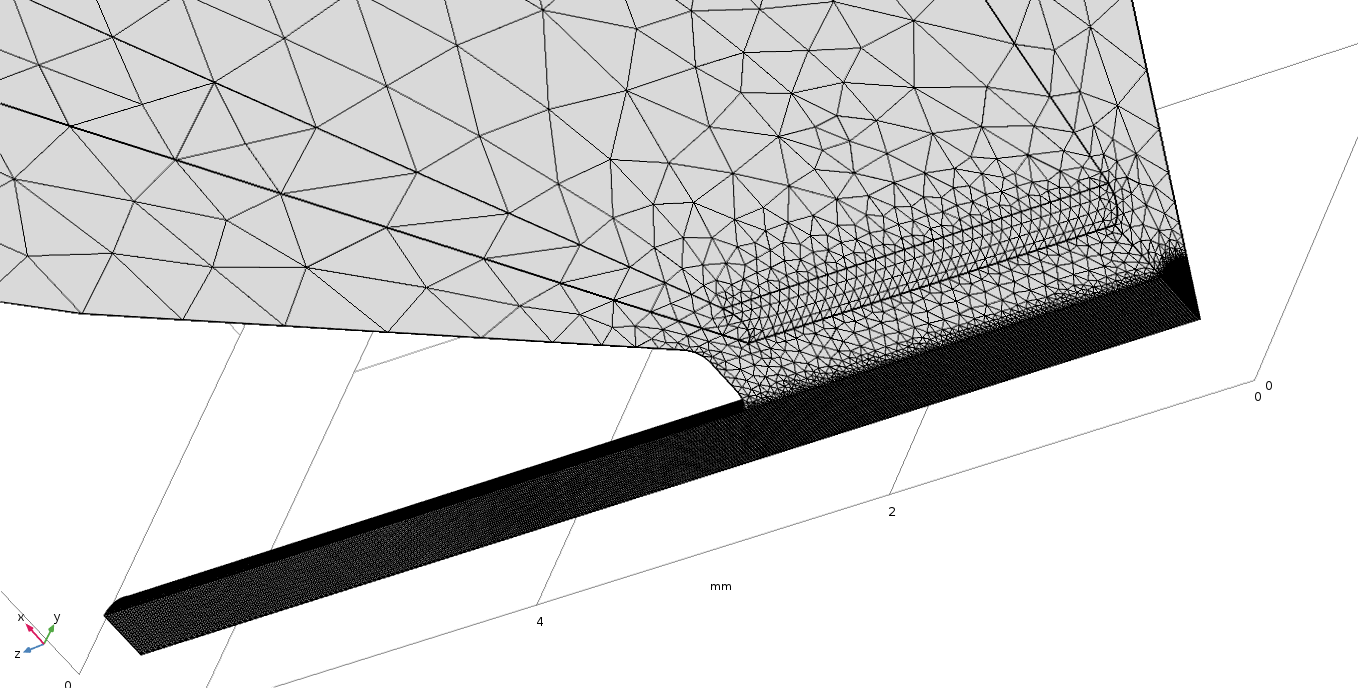}\hspace*{-5mm}
\includegraphics[scale=0.15]{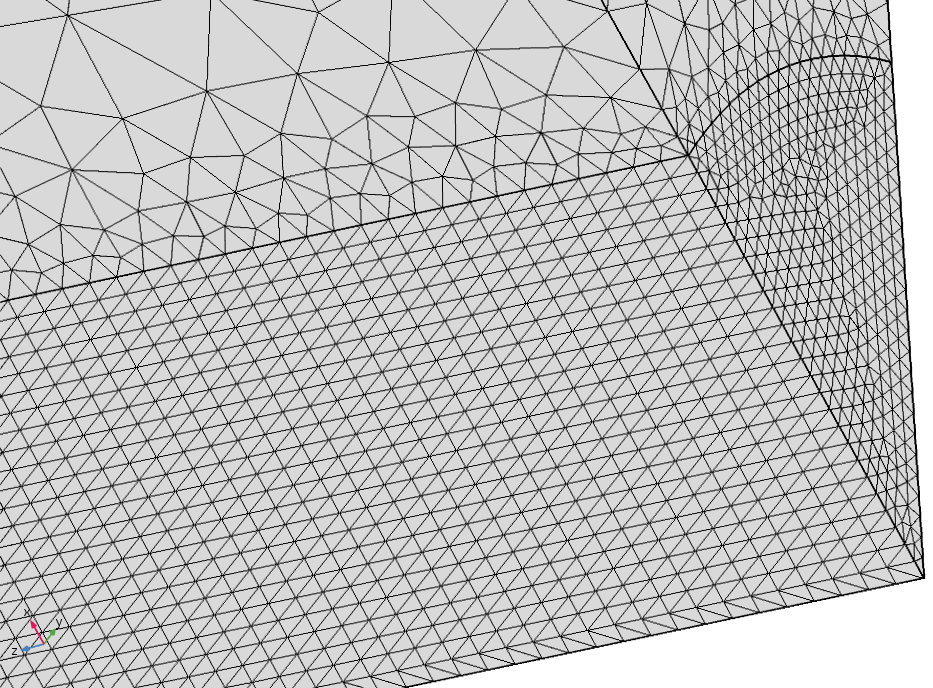}
\caption{FEM mesh setup in the simulated segment of the trap field (only 1/16 of the entire geometry necessary due to symmetries and antisymmetries). Only the vacuum part in the critical region of the trap needs to be meshed. The two closer views show the finer mesh around curved electrode surfaces, the mesh in the wedge adjacent to the axis shows the arangement of the mesh points in planes perpendicular to the axis.}
\label{meshsetup}
\end{figure}

We simulated the electric field in Comsol version 4.4 using the stationary solver.
The FEM shape functions were quadratic.
Fig. \ref{meshsetup} shows the meshed part of the geometry

\begin{figure}
\begin{tabular}{cc}
$P_{++}$ & $P_{+-}$ \\
\includegraphics[width=.47\textwidth]{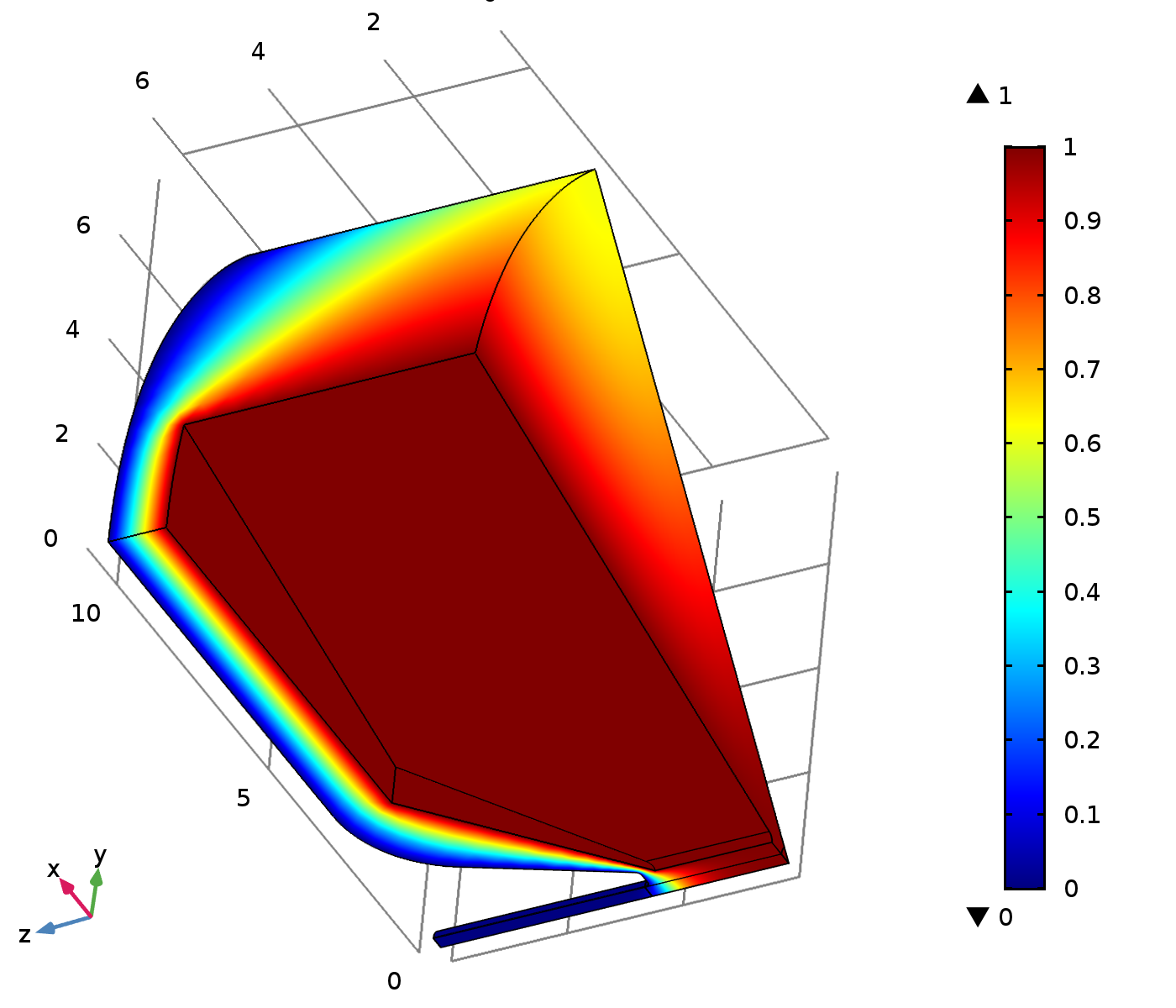} & \includegraphics[width=.47\textwidth]{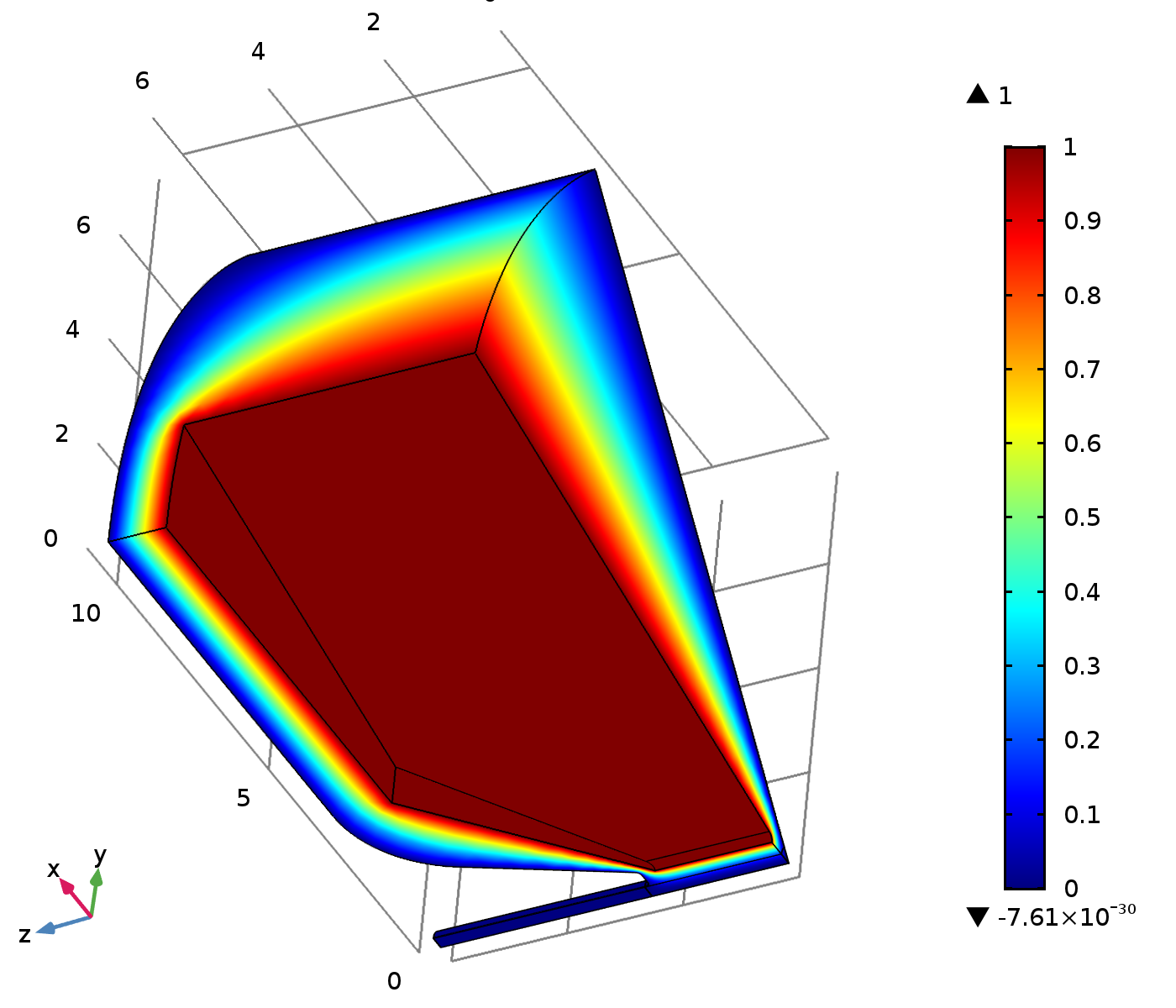} \\
\includegraphics[width=.47\textwidth]{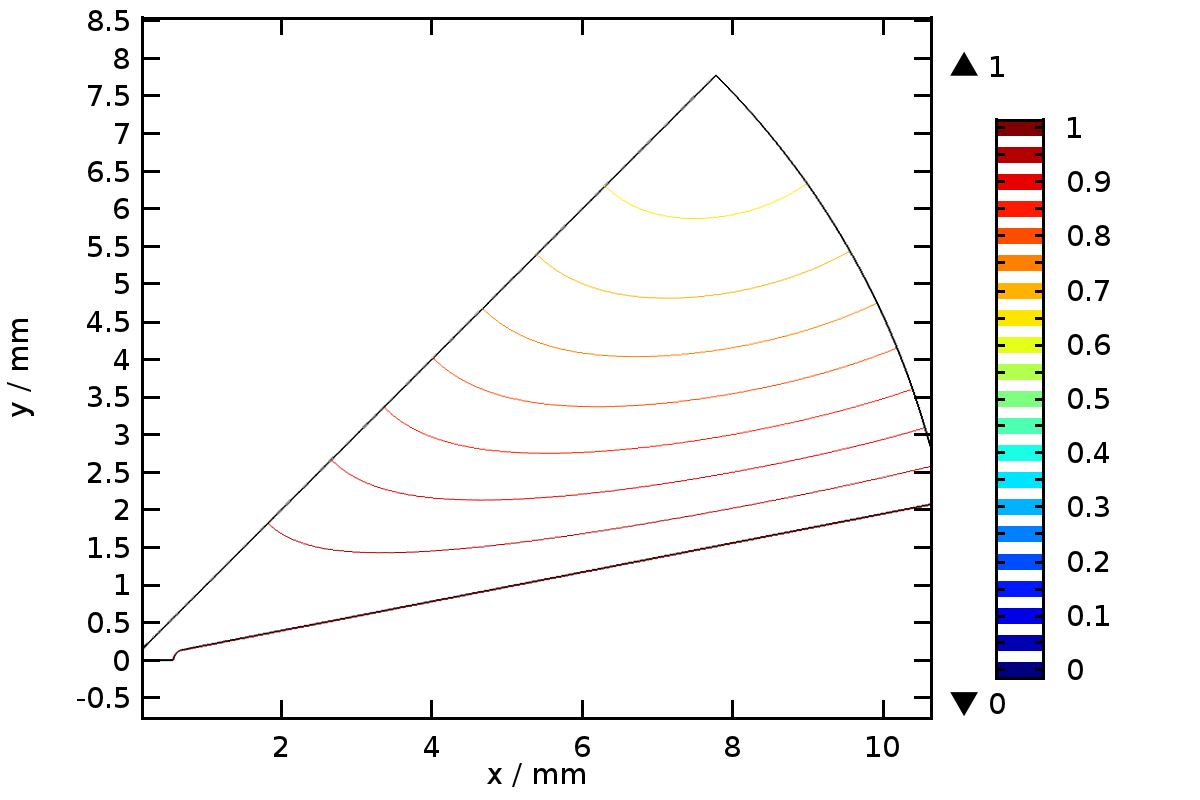}  & \includegraphics[width=.47\textwidth]{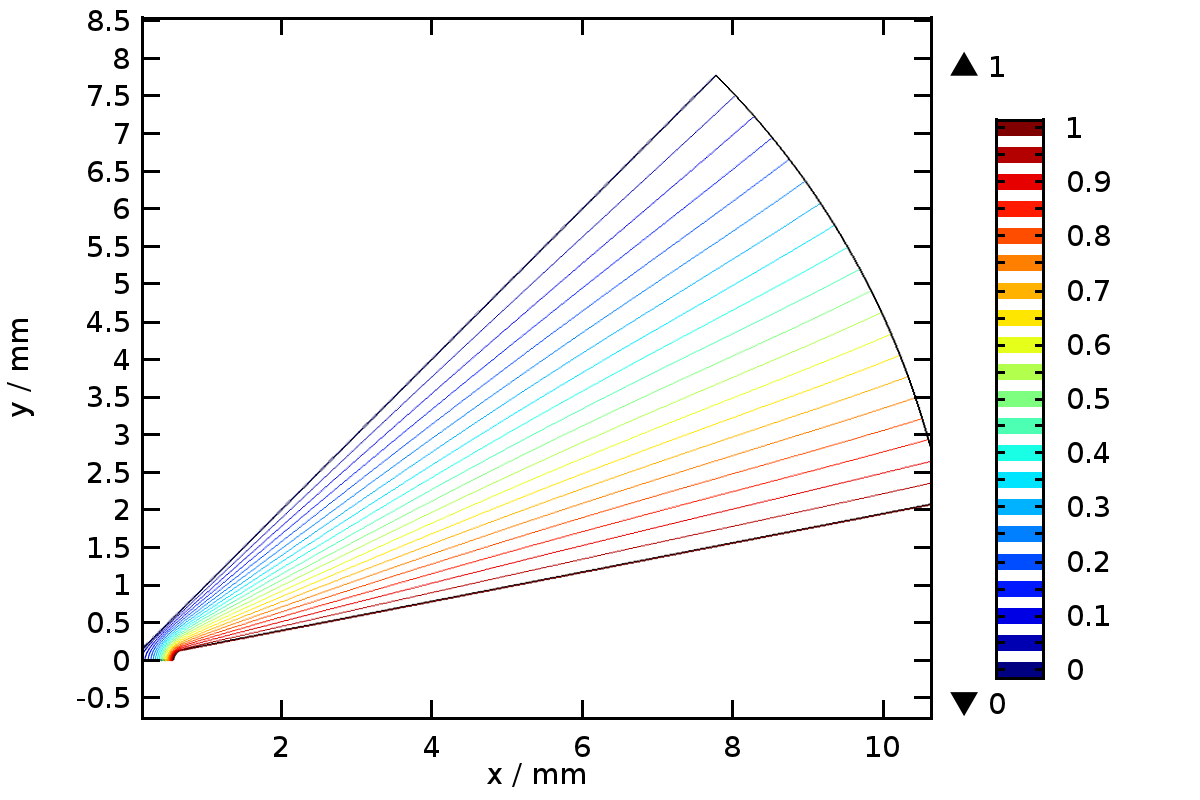} \\
\includegraphics[width=.47\textwidth]{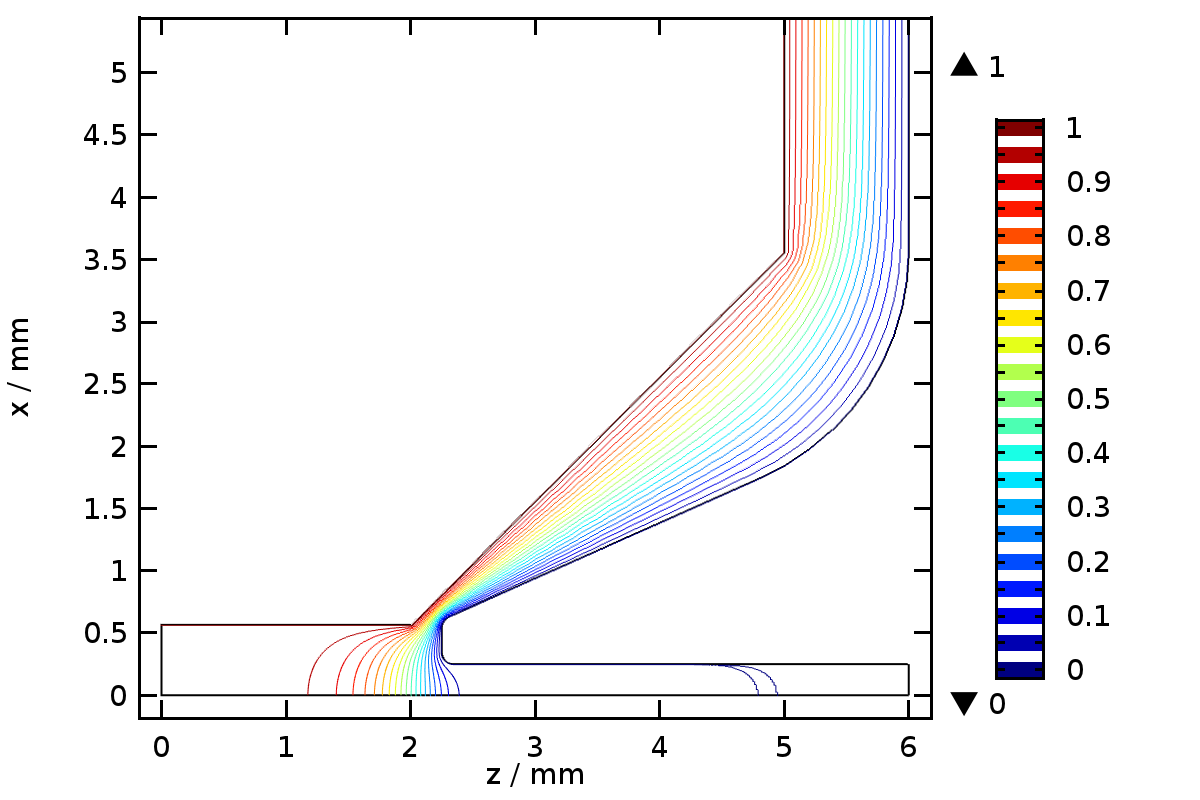}  & \includegraphics[width=.47\textwidth]{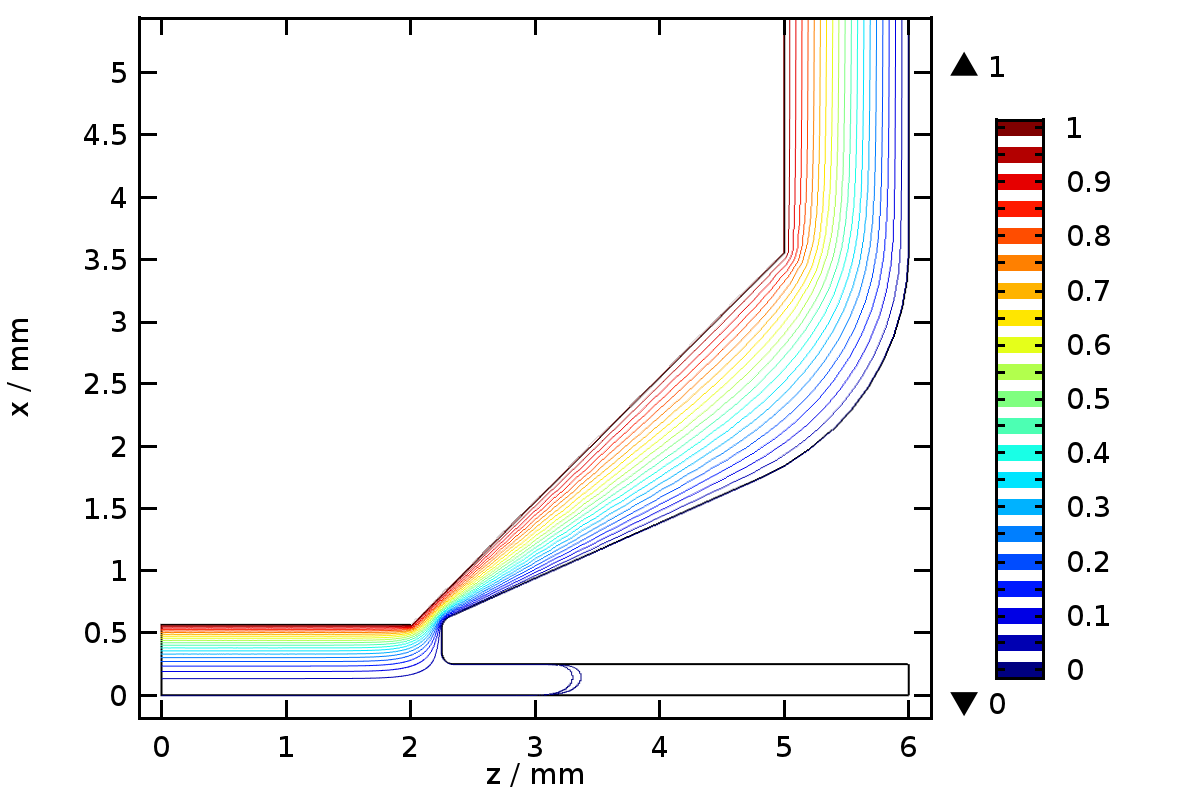} \\
\end{tabular}
\caption{Results of the FEM calculation of the two basis distributions in eq. (\ref{linkom}).
$P_{++}$ corresponds to grounded endcap electrodes and all the quadrupole electrodes at a unit potential, there is the symmetry condition on the surface $x=y$.
$P_{+-}$, having the antisymetry condition (a zero equipotential) in the plane $x=y$, represents the quadrupole electrodes at alternating unit and negative unit potentials with the edncap electrodes grounded.
Both basis functions are symmetrical about the planes $y=0$, $x=0$ and $z=0$.}
\label{femresults}
\end{figure}

The same FEM mesh was used for both potential basis functions $P_{++}$ and $P_{+-}$, they only differed in boundary conditions, described in section \ref{potdecomp}.
The Dirichlet boundary conditions define fixed potentials on the electrode surfaces, a zero equipotential on a boundary presents the antisymmetry condition on it.
The Neumann boundary condition (zero normal component of the electric field) then introduces the symmetry condition on that surface.
The type of the condition on each surface of the simulated segment can be seen in the plots of the calculated results in Fig.~\ref{femresults}.

\subsection{Extraction of the Multipole Axial Functions}
We fitted the expansions of the electric potential (\ref{ex++rp}) and (\ref{ex+-rp}) with terms of orders up to $r^6$ on the data from the FEM calculation for the basis functions $P_{++}$ and $P_{+-}$, respectively, using the procedure described in section \ref{multextr}.
Only the mesh points within the distance $r_{max}=0.14\,mm$ from the axis were used in the regression analysis.
That distance was selected based on the trade-off between the radius of validity of the analytical formulas and the attainable standard deviation of the calculated values of the axial functions $p_{\mu,n}$.
To accommodate the changes of the potential further from the axis, the model functions (\ref{ex++rp}) and (\ref{ex+-rp}) would have to contain higher order terms than $r^6$.
The value $r_{max}$ then also defines the approximate region of validity of the field interpolation.

\begin{figure}
\includegraphics[width=\textwidth]{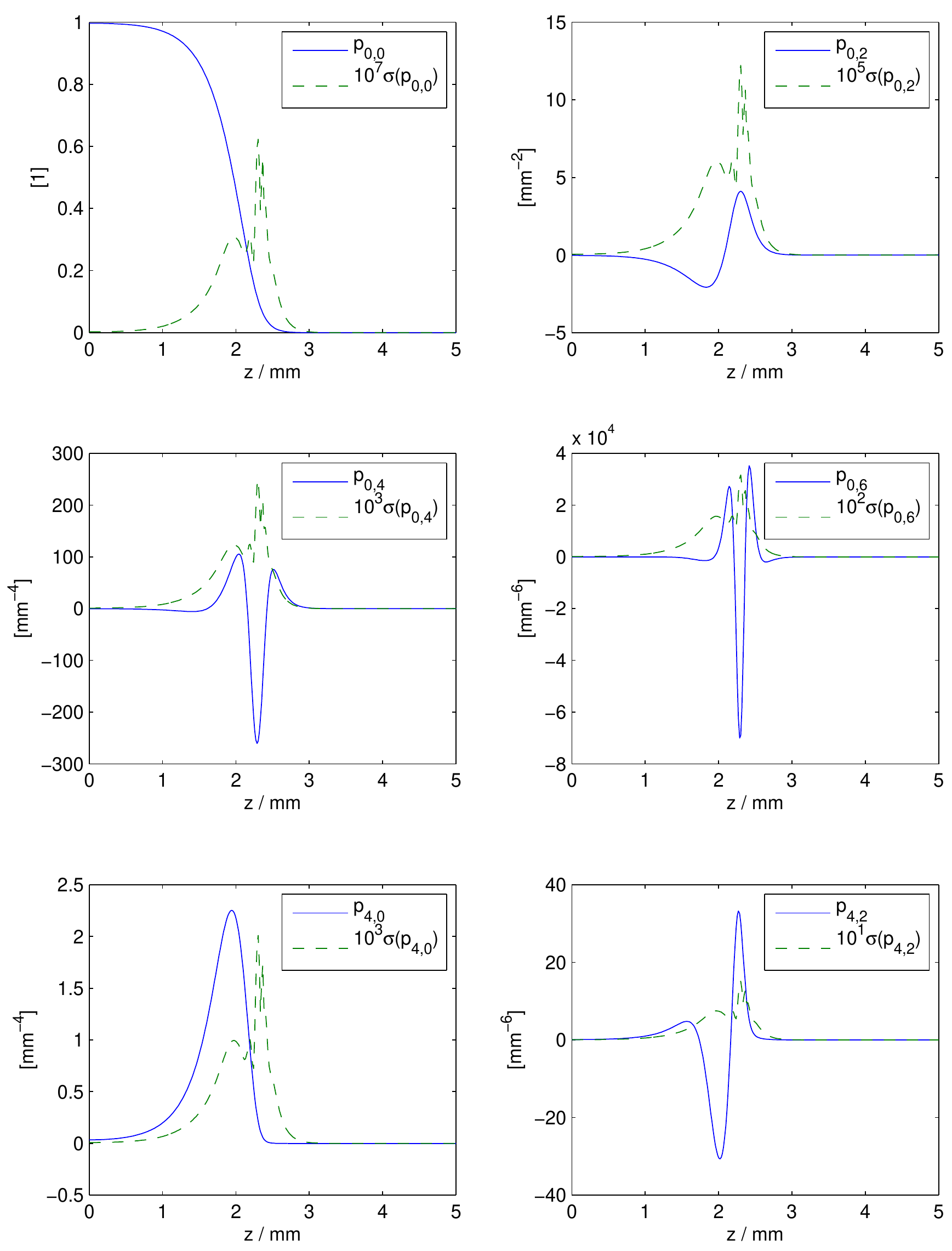}
\caption{Axial multipole functions in the expansion of $P_{++}$ (\ref{ex++rp}), $\sigma(p_{\mu,n})$ are the corresponding statistical estimates of the standard deviation. The curves are symmetrical about $z=0$. }
\label{fitP++}
\end{figure}

\begin{figure}
\includegraphics[width=\textwidth]{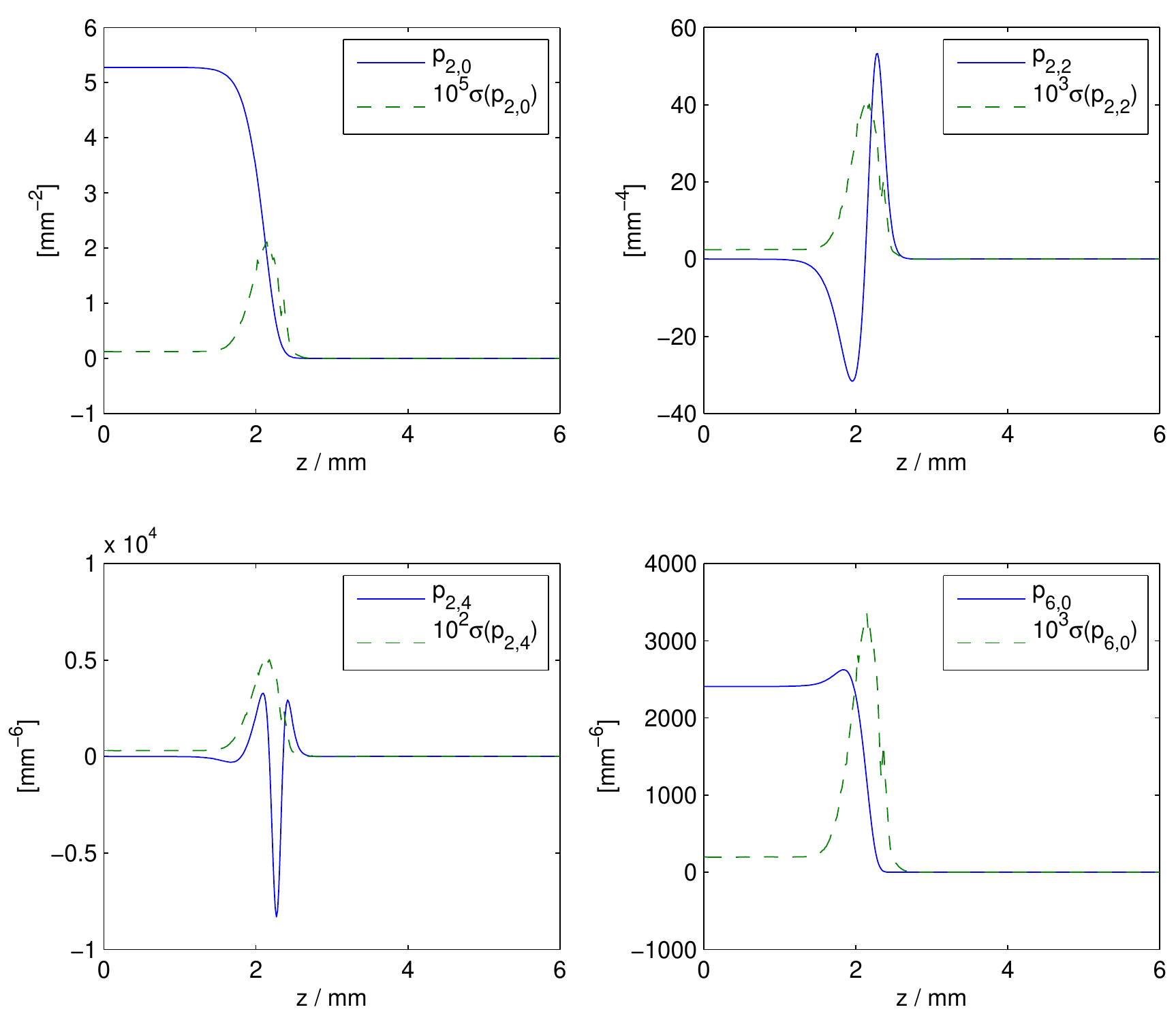}
\caption{Axial multipole functions in the expansion of $P_{+-}$ (\ref{ex++rp}), $\sigma(p_{\mu,n})$ are the corresponding statistical estimates of the standard deviation. The curves are symmetrical about $z=0$.}
\label{fitP+-}
\end{figure}

The results plotted in Figs. (\ref{fitP++}) and (\ref{fitP+-}).
The values of the standard deviation estimates indicate that the key axial functions were determined with a high accuracy.

\subsection{Results and discussion}
We have selected several results from the trajectory calculation code that demonstrate some interesting features of ion trajectories in the presented trap under various conditions.

\begin{figure}
\includegraphics[width=\textwidth]{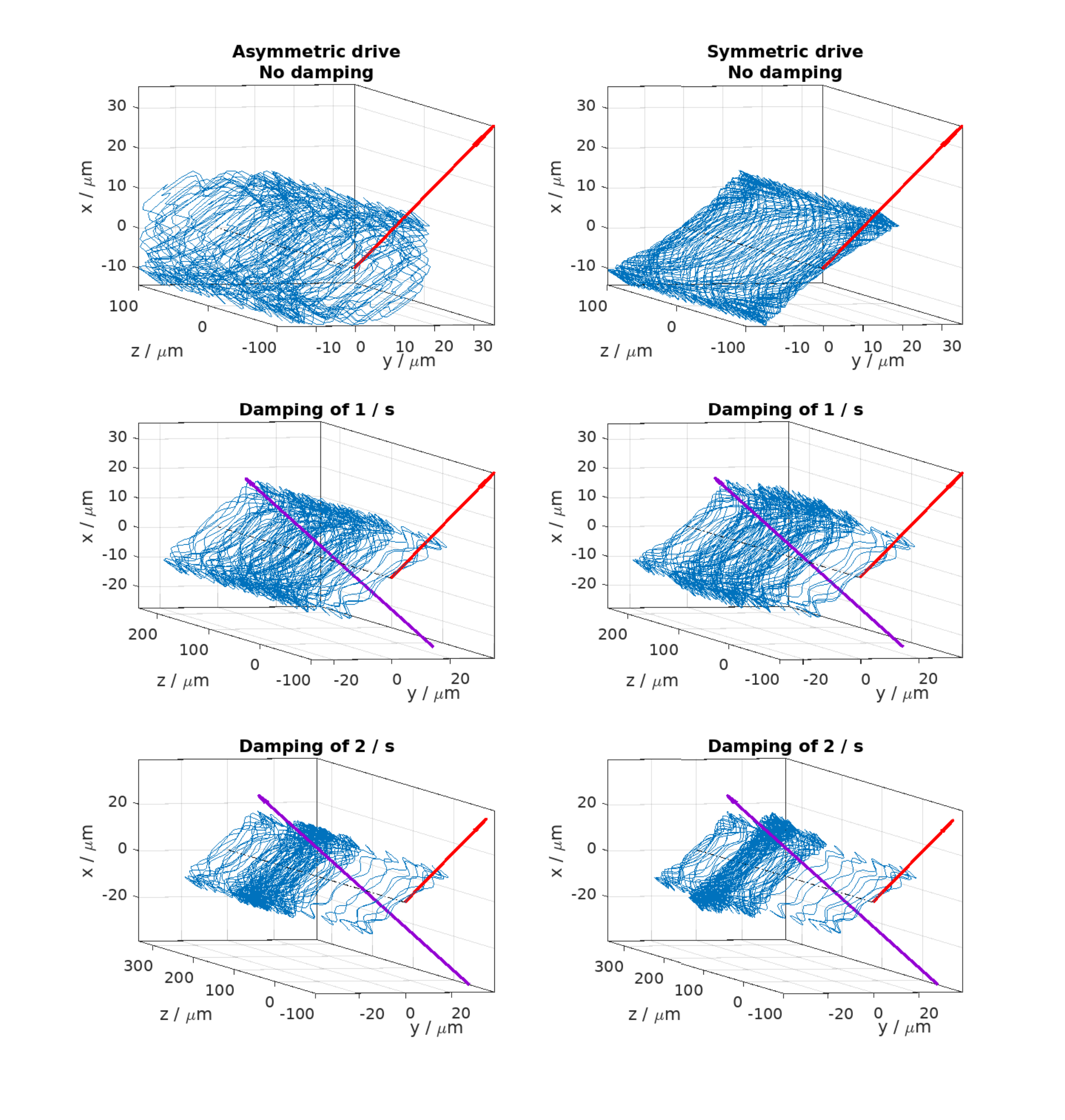}
\caption{Single ion trajectories for the two wiring schemes (on in each column of the trap and different damping coefficients $f$.
The red arrow on the left of each plot shows the direction of the initial velocity, its magnitude $v=\sqrt{kT/m}$ corresponds to the thermodynamic temperature of 800\,K.
The purple line at the origin of the coordinate system shows the direction of the cooling laser beam, the vector with Carthesian components $(1,-1,1)/\sqrt{3}$.
A unit ratio of the scales along the axes in each plot.}
\label{res-damping}
\end{figure}

Fig. \ref{res-damping} shows a single Ca$^{+}$ ion with the same initial position and velocity in a trap driven asymmetrically and symmetrically with $U_{AC}=600$\,V and $U_{DC}=10$\,V.
One can identify the micromotion superimposed onto the secular oscillations.
With no damping, the particle would oscillate in the trap indefinitely.
The symmetric setup preserves the plane of the initial velocity in the projection into the $xy$-plane.
The damping lowers the amplitude of the secular oscillations, ultimately bringing the ion to rest at the center of the trap.
A higher damping factor naturally leads to a faster relaxation but that has its limits.
The symmetric wiring provides a stiffer ponderomotive force, which leads to faster secular oscillations with a lower amplitude.

Increasing the damping leads to a longer range of motiong along the $z$-axis.
That can be explained by the following two effects: Since the damping acts in both axial and transversal directions, given the nature of the force (\ref{dragforce}), the motion in the transversal direction is partially transferred into the axial motion and vice-versa. Additionally, an increased damping also limits the micromotion -- the principal feature of the trajectory which holds the ions in the trap.
The confining effect of the trap is then weakened by the damping.
If the damping coefficient is further increased, the ion will escape the trap.

The unidirectional damping force gradually eliminates any motion in its direction, so that after some time the motion tends to stay in the direction perpendicular to the cooling beam.
This largely happens in the projection in the $xy$ plane, due to the trap's elongated shape.
However, this is more clearly visible in the symmetric setup only, as the stiffness for the secular motion in the $xy$ projection is uniform in all directions, and the particle is thus allowed to only retain motion perpendicularly to the damping beam in that direction.
On the other hand, the asymmetric driving, the secular oscillations along the $x$-axis and along the $y$-axis are coupled -- the motion may be dampened in the direction of the cooling beam, but the motion in the remaining degrees of freedom will gradually reappear in the dampened direction again, eventually allowing the ion to be cooled down in all degrees of freedom.

\begin{figure}
\begin{tabular}{cc}
{\footnotesize\sf Asymmetric drive} & {\footnotesize\sf Symmetric drive} \\
\noindent\parbox[c]{.47\hsize}{\includegraphics[width=.48\textwidth]{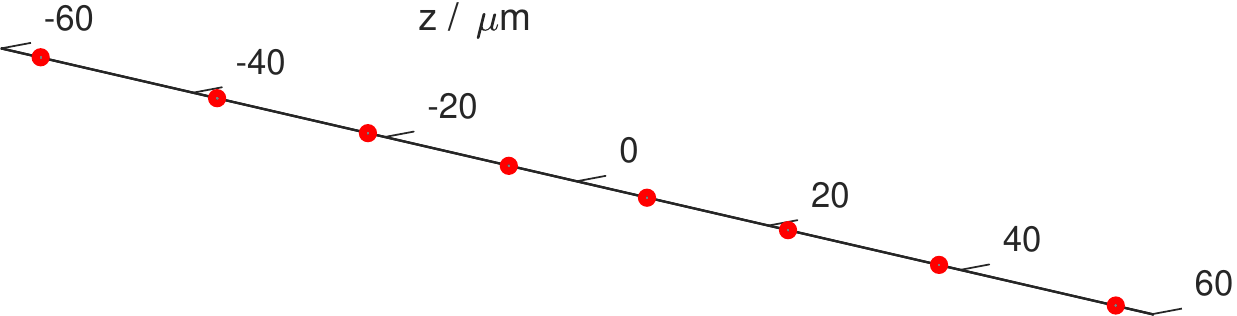}} &
\noindent\parbox[c]{.47\hsize}{\includegraphics[width=.48\textwidth]{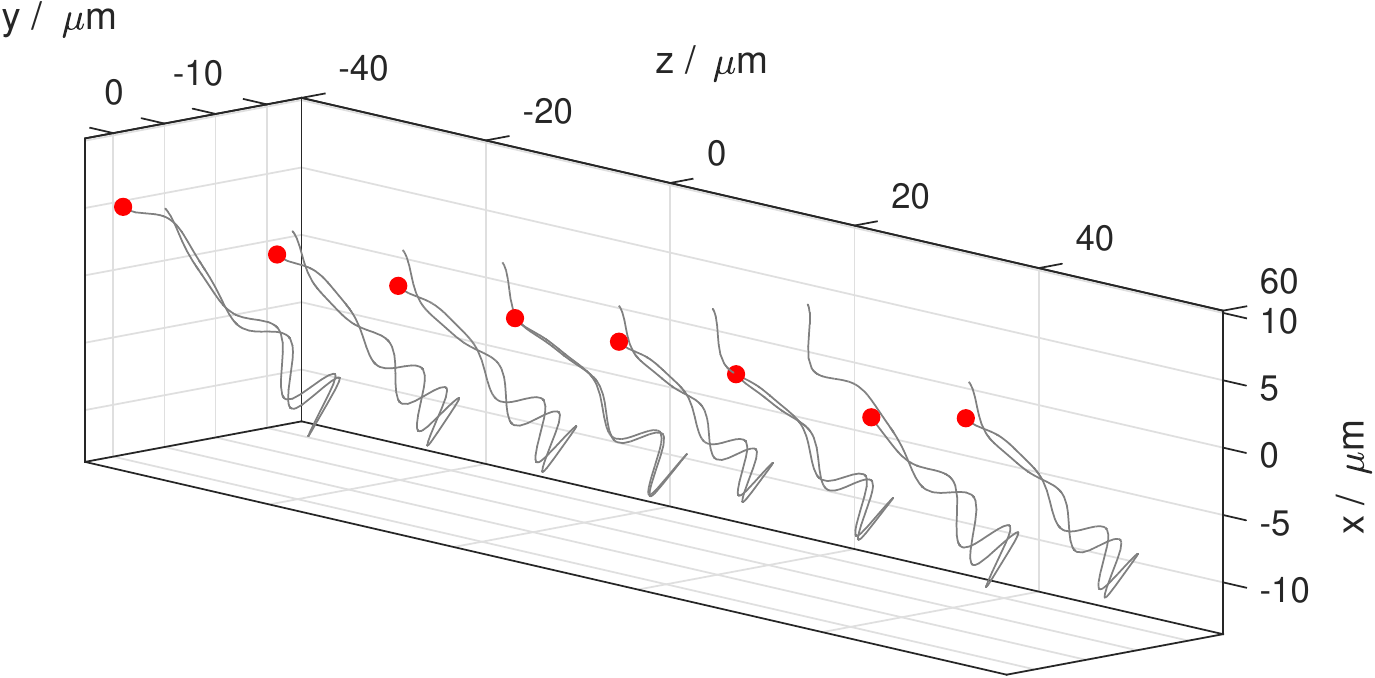}}
\end{tabular}
\caption{The effect of unidirectional damping in the direction of $(1,-1,1)/\sqrt{3}$ on a chain of eight Ca$^{+}$ ions. The grey curly lines are the ion trajectories in the 500 integration steps taken prior to the positions marked with the dots. In the asymmetrically driven trap all degrees of freedom are damped, in the
symmetrical case the motion perpendicular to the direction of damping remains.}
\label{res-8-sym+asym}
\end{figure}

That is also evident in the motion of multiple ions.
In Fig. \ref{res-8-sym+asym}, the motion along the $z$-axis is dampened in both driving types, as the damping vector has a non-zero $z$-component. In the symmetric driving setup, the ions still retain motion perpendicularly to the cooling direction in the $xy$ projection, while with the asymmetric driving, the motion is eliminated entirely, and the ions assume positions given by the balance between their mutual repulsive Coulomb forces and the field of the trap along the $z$ axis generated by the endcap electrodes at a positive potential.

\section{Conclusions}
We calculated the electric field in a linear quadrupole Paul trap using the second-order FEM in Comsol.
To increase the accuracy of the FEM calculation, and to decrease the computational effort, we utilize symmetries and antisymmetries exhibited by the trap's geometry and the potential on the electrodes.
From the the FEM data we extracted the coefficients in the azimuthal multipole potential expansion along the main axis of the trap.
Besides the essential multiple components, the quadrupole and the rotationally symmetric part, we also included higher multipoles, as they were determined with a good accuracy from the FEM data.

We use the multiple expansion to interpolate the electric field components in a simulation of ion trajectories in the trap.
The simulation code allows for the Coulomb repulsion for more than one ion.
We have also included a model for Doppler cooling in a form a unidirectional damping force with the magnitude proportional to the projection of the ion's velocity into the direction of the cooling beam.

We show the simulation code on selected examples of ion trajectories for two practical wiring schemes of the trap: the asymmetric setup, in which only one pair of the quadrupole electrodes is driven at a time dependent voltage, and the symemtric one, in which one pair is driven in a counterphase to the other pair.
Our results show that the asymmetric driving allows to slow down ions by damping in a single direction as much as possible due to all degrees of freedom being coupled.
Then, the only remaining motion is the micromotion.

\section*{Acknowledgements}
Parts of this research were performed within the EMPIR project 17FUN07 CC4C. The financial support of the EMPIR initiative is gratefully acknowledged. The EMPIR initiative is co-funded by the European Union's Horizon 2020 research and innovation programme and EMPIR Participating States.

We acknowledge the kind technological support from the group of Rainer Blatt from Universtit\"at Innsbruck including the contribution of Yves Colombe and Kirill Lakhmanskiy to the construction of the employed linear Paul trap.

The research was supported by the TA CR (TE01020118), the MEYS CR (LO1212), its infrastructure by the MEYS CR and the EC (CZ.1.05/2.1.0 0/01.0 017) and by the CAS (RVO:68081731).

\end{document}